\documentclass[epj]{svjour}
\usepackage{graphics,wrapfig}

\begin{document}

\newcommand{\qs}{$Q^2$~}
\newcommand{\qsat}{$Q_S^A$~}
\newcommand{\x}{$x$~}
\newcommand{\ea}{$e$+A~}
\newcommand{\ep}{$e$+$p$~}
\newcommand{\pom}{I\!\!P}

\title{Hot Quarks and Gluons at an Electron-Ion Collider}
\author{Matthew A. C. Lamont\inst{1} for the EIC Collaboration
}                     
\offprints{}          
\institute{Brookhaven National Laboratory, Upton, NY 11973 USA}
\date{Received: date / Revised version: date}
%
\abstract{
The nuclear wave-function is dominated at low- and medium-x by gluons.  As the rapid growth of the gluon distribution towards low $x$, as derived from current theoretical estimates, would violate unitarity, there must be a mechanism that tames this explosive growth.  This is most efficiently studied in colliders running in $e$+A mode, as the nucleus is an efficient amplifier of saturation effects occurring with high gluon densities.  In fact, large A can lead to these effects manifesting themselves at energies a few orders of magnitude lower than in $e$+$p$ collisions.  In order to study these effects, there are proposals to build an $e$+A machine in the USA, operating over a large range of masses and energies.  These studies will allow for an in-depth comparison to A+A collisions where results have given tantalising hints of a new state of matter with partonic degrees of freedom.  In order to explain these results quantitively, the gluons and their interactions must be understood fully as they are the dominant source of hard probes at both RHIC and LHC energies.
} 
\maketitle

\section{Understanding the Gluonic Space-Momentum Distributions in Nuclei}

Although all of the unique features of QCD are determined by the self-interactions of gluons (such as Asymptotic Freedom), little is known about their space- and momentum-distributions in both nucleons and nuclei.  Gluons are the mediators of the strong interaction, they dominate the structure of the QCD vacuum, yet a study of gluon properties is difficult as the gluonic degrees of freedom are missing in the hadronic spectra.  Therefore, high-energy probes of the nucleus are required.  Whilst $p$+A collisions provide excellent information on the gluon properties (as many observables require gluons to participate at the leading order), soft colour interactions between the $p$ and the A both before and after the hard scattering mean that interpreting the data is difficult, if not impossible.

Therefore, the most desirable system with which to probe the gluon properties are with lepton+A collisions, which are dominated by single-photon exchange and hence have a better chance to preserve the properties of the partons in the nuclear wave function.  Although using an electromagnetic process to study gluons may sound counter-intuitive, it is, in fact, the most precise tool available.  The lepton beams interact with the electrically charged quarks in a process known as Deep-Inelastic Scattering (DIS) and the gluonic part of the nuclear wave-function modifies the well-understood interaction in ways which allow for the extraction of the gluon properties.

\begin{figure}
\resizebox{0.50\textwidth}{!}{%
   \includegraphics{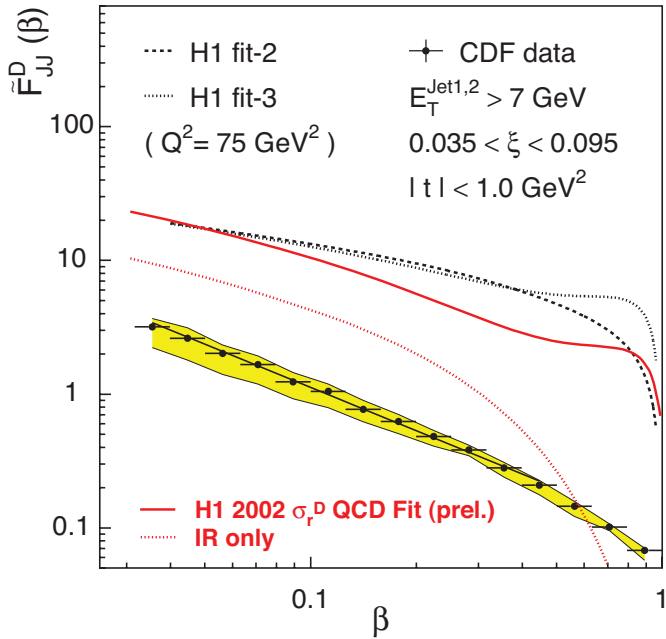}
}
\caption{A comparison of diffractive di-jet data from CDF to a prediction based on diffractive parton distribution functions (dpdfs), assuming factorization to hold (taken from \cite{Ref:pAvseA}).}
\label{Fig:pAvseA}
\end{figure}

The importance of the additional effects in $p$+A collisions is best outlined in Figure ~\ref{Fig:pAvseA} (taken from \cite{Ref:pAvseA}).  Here, diffractive parton distribution functions were obtained from data at HERA ($e+p$ collisions).  These were then used to predict diffractive di-jet production at the Tevatron ($p+\overline{p}$) and these predictions are compared to actual CDF data.  As can be seen, there is up to an order of magnitude difference and more between the prediction and the data, indicating a complete breakdown of factorization at hadron-hadron colliders.

The invariant cross-section in DIS in next-to-leading order can be written as:

\begin{eqnarray*}
&& \frac{d^2 \sigma^{eA \rightarrow eX}}{dx dQ^2} = \\ 
&& \frac{4 \pi \alpha^2_{e.m.}}{xQ^4} \left[ \left(1-y+\frac{y^2}{2} \right )
    F^A_2(x,Q^2) - \frac{y^2}{2} F^A_L(x,Q^2) \right]
\label{Eqn:DIS}
\end{eqnarray*}

where $y$, the inelasticity, is the fraction of the energy lost by the lepton in the rest frame of the nuclei.  $F^A_2$ represents the quark and anti-quark structure function and the longitudinal structure function, $F^A_L$, represents that of the gluons.  The dependence of $F_2$ on $x$ was studied extensively at HERA for nucleons, where the gluonic properties were inferred through the scaling violation of this structure function.  In order to determine $F_L$ directly, measurements must be made at different energies.  This was performed in the final years running at HERA and first results are currently appearing in the literature~\cite{Ref:HERA_FL}.

\subsection{Gluon Saturation}

\begin{figure}
\resizebox{0.50\textwidth}{!}{%
   \includegraphics{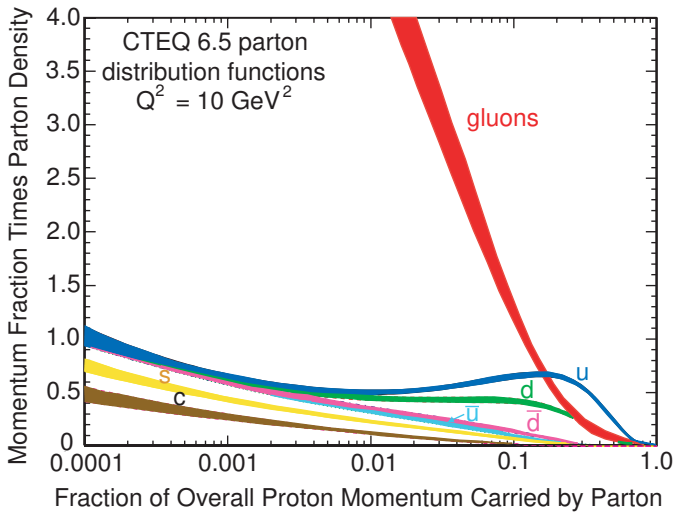}
}
\caption{The gluon and quark momentum distributions in the nucleon as a function of $x$, the fraction of the overall proton momentum carried by the parton, using the CTEQ 6.5 parton distribution functions available online~\cite{Ref:Durham}. (Plot taken from \cite{Ref:Vigdor}).}
\label{Fig:GluonSat}
\end{figure}

As shown in Figure \ref{Fig:GluonSat}, a well known phenomenon observed in DIS experiments on protons at HERA is that for $Q^2 \gg \Lambda^2$, the gluon density in the nucleon increases rapidly as $x$ decreases and is significantly greater than that of the quarks for $x <$ 0.01.

\begin{figure*}
\resizebox{1.0\textwidth}{!}{%
    \includegraphics{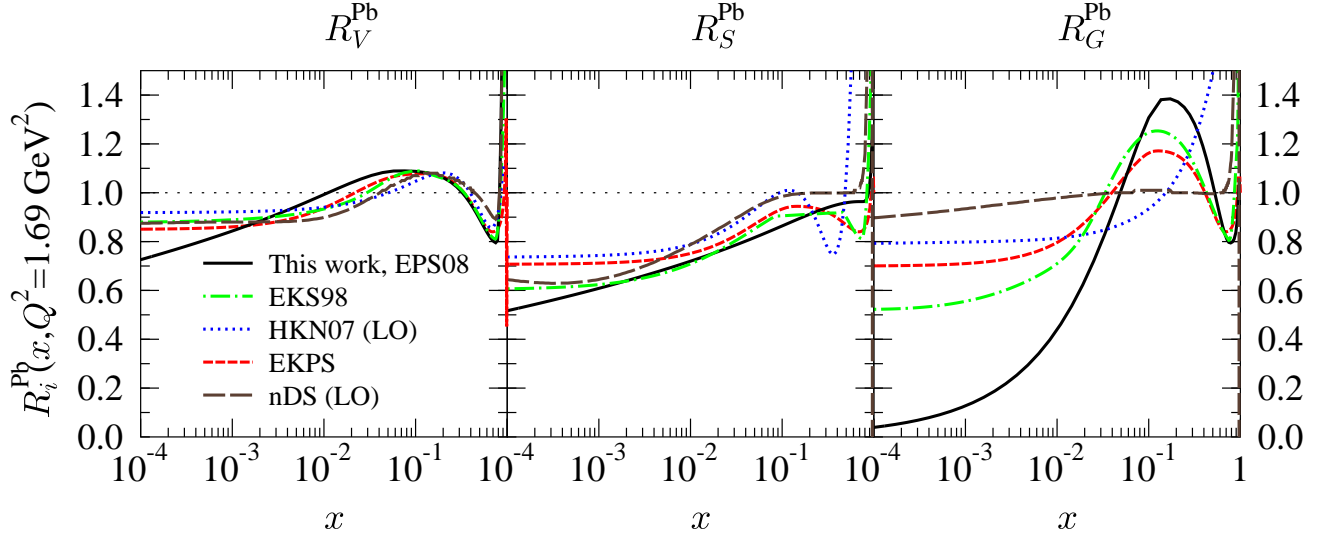}
}
     \caption{A comparison of the average valence, sea-quark, and gluon modiÞcations at $Q^2$ = 1.69 GeV$^2$ for the Pb nucleus for different LO global DGLAP analyses.  Please consult the original reference for a more detailed discussion of the different models in the plot~\cite{Ref:Salgado}}
    \label{Fig:GluonDists}
\end{figure*}

In DIS experiments at nuclei at small $x$, it is found that the quark and gluon distributions are modified when compared to their distributions in nucleons, an effect known as shadowing.  At smaller $x$ ($<$ 0.01), there are currently no existing measurements of the gluon distributions.   The importance of providing constraints on this measurement is outlined in Figure~\ref{Fig:GluonDists}.  This shows a comparison of nuclear effects (for a Pb nucleus at a fixed $Q^2$) in the average valence-quark, sea-quark and gluon distributions for various LO linear DGLAP analyses (for details, refer to~\cite{Ref:Salgado}).  It can immediately be seen that there is very good agreement between the various models for the valence- and sea-quark modifications, but large differences between the models for the case of the gluons, especially so for the EPS08 model which takes into account the data at forward rapidities from the BRAHMS experiment at RHIC~\cite{Ref:BRAHMS}.  There are significant differences between the models already in the ``forward RHIC rapidity" region (\x $\sim$ 10$^{-3}$) and these are even greater at lower $x$.  This is particularly important for heavy-ion collisions at the LHC where the mid-rapidity region will occur at \x $\sim$ 10$^{-4}$.  Therefore, a full knowledge of the gluon properties as a function of \x is required to enable the interpretation of the LHC data.

At large $x$ and $Q^2$, the gluon properties are determined theoretically by linear evolution equations (DGLAP~\cite{Ref:DGLAP} along \qs and BFKL~\cite{Ref:BFKL} along $x$).  The rapid rise of the gluon densities at low-\x is believed to arise from gluon Bremstrahlung - that is, hard (high-$x$) gluons shed successively softer (low-$x$) gluons.  At small enough \x, in order to avoid violating unitarity constraints, gluon saturation must occur.  That is, the gluon Bremstrahlung process is matched by low-\x gluons recombining to form harder gluons.  The high number of gluons involved in these processes means that the dynamics are classical and their piling up at the saturation scale, \qsat, is reminiscent of a Bose-Einstein condensate.  This has lead to suggestions that the gluonic matter in nuclear wave-functions at high energies is universal and can be described as a Colour Glass Condensate~\cite{Ref:CGC}.

\subsubsection{The Nuclear ``Oomph Factor"}

\begin{figure}
\resizebox{0.5\textwidth}{!}{%
   \includegraphics{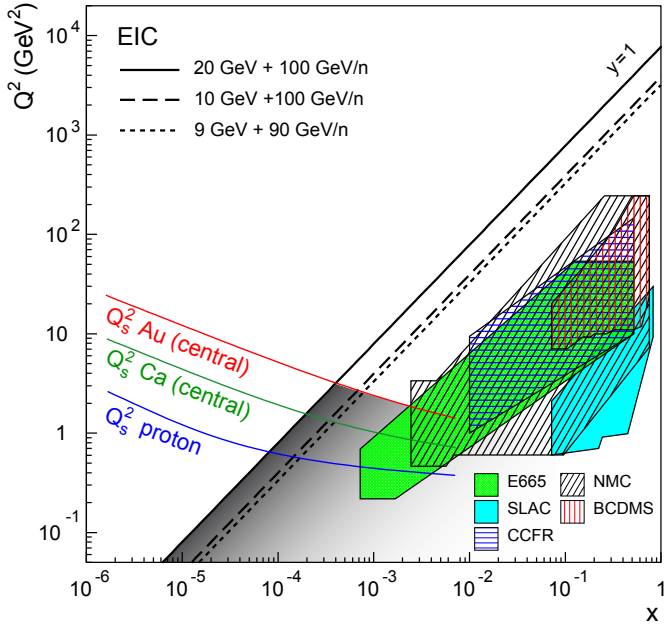}
}
\caption{The kinematic acceptance in \x and \qs for different energies in existing $l$+A systems and a proposed Electron-Ion Collider.  The \qs scale is shown for different A species.}
\label{Fig:l+A}
\end{figure}

This process is described by the non-linear, small-\x JIMWLK renormalization group equations ~\cite{Ref:JIMWLK}.  The onset of saturation described by the JIMWLK equations is characterised by a dynamical saturation scale, $Q_S^2$.  This scale grows with smaller \x (or larger energy) and increasing nuclear size, meaning that it is experimentally more accessible in heavy \ea collisions than in \ep.  Simple geometrical considerations suggest that the saturation scale, $Q_s^2 \propto (A/x)^{1/3}$.  This means that the nucleus acts as an efficient amplifier of the physics of high gluon densities.  More rigourous calculations support this outcome, with the dependence on A even larger than 1/3 for high A~\cite{Ref:Oomph}.  As the authors note, it is important to consider the correct density profile of the nucleon when calculating the Oomph factor - that is, the mean impact parameter is not at $b=0$, but rather at $b$=0.4 $fm$.  For large A, there is a significant region at low \x where $Q_s^2 \gg Q^2 \gg\Lambda_{\rm QCD}^2$ and the regime of strong non-linear fields is applicable.  The saturation regimes in $x-Q^2$ space are shown in Figure \ref{Fig:l+A} for different ion species.  As is evident from the figure, this region has not yet been accessible to previous $\ell$+A collision systems but will be accessible by a proposed Electron-Ion Collider, which, at the proposed energies and luminosities will provide for data well into the saturation regime according to the latest theories.  This new accelerator proposal will be discussed further in section \ref{Sec:EIC}.

\section{Experimental Observables}

The key questions which define the experimental observables in \ea physics can be grouped into the following categories:

\subsection{\textbf{What are the momentum distributions of gluons and sea-quarks in nuclei?}}

The knowledge of the momentum distributions of partons in nuclei is one of the first key measurements to be made at an EIC.  There are a number of ways of extracting $xG^A(x, Q^2)$.  This can be achieved through: $i)$ measuring the scaling violation of $F_2^A$ (the quark structure function) with \qs ($\partial F_2^A/\partial \ln(Q^2) \neq 0$).  This method was used successfully as HERA but is not in and of itself a direct measurement.  $ii)$ a direct measurement of $F_L^A$.  This is the preferred method and a quick examination of equation \ref{Eqn:DIS} shows that its dependence on $y$ means that this can only be accomplished through running at more than one energy (as $y$ = $Q^2/xs$).  This has also been performed in the last year of data taking with HERA, allowing for a small-statistics direct measurement of $F_L$ of the proton.  $iii)$ the measurement of inelastic and $iv)$ diffractive vector meson production.

One notable characteristic of an EIC will be the ability to measure the $F_s^C$ and $F_L^C$ charm structure functions.  These are important as they are sensitive to the photon-gluon fusion process at high energies and no data currently exists on nuclear charm-quark distributions for \x $<$ 0.1.  The high luminosities of a proposed EIC will lead to the production of 10$^5$ charm pairs for 5 fb$^{-1}$ of data, enabling precision studies.

\subsection{\textbf{What are the space-time distributions of gluons and sea-quarks in nuclei?}}

As well as understanding correctly the momentum distributions of the gluons and sea-quarks in nuclei, it is also desirable to measure their space-time distribution (i.e. the gluon density profile) in order to understand the physics of high gluon densities.  Very little is currently known about this (such as, is the glue distributed uniformly or is it ``clumpy"?).

In order to extract this information, it is worthwhile to think of DIS physics at small-\x as taking place in a frame where the virtual photon fluctuates into a quark-anti-quark dipole.  This dipole then subsequently scatters coherently on the nucleus (or hadron).  By measuring the vector mesons which evolve out of the $q\overline{q}$ pair, then taking the Fourier transform of the vector meson cross-section as a function of the momentum transfer between the initial and final-state hadrons/nuclei, one can estimate the scattering matrix for this amplitude.  Using the optical theorem, one can extract the $survival$ $probability$ of small-sized dipoles ($d$ $\ll$ 1 fm) to propagate through the target at a given impact parameter without interacting.

\begin{figure}
\resizebox{0.5\textwidth}{!}{%
   \includegraphics{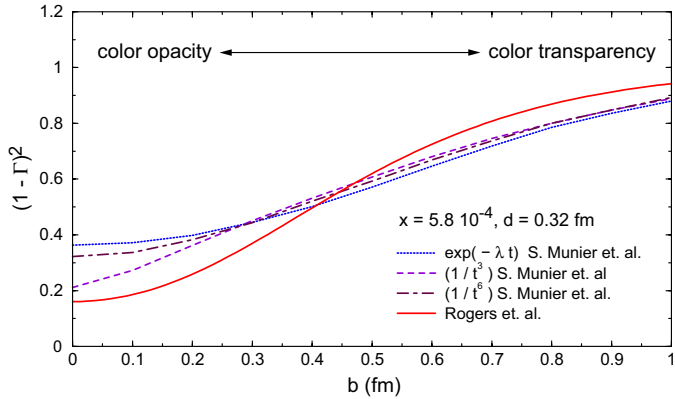}
}
\caption{The survival probability of vector mesons as a function of impact parameter, extracted from HERA data on elastic vector meson production on the proton.  The Munier et al. curves \cite{Ref:Munier} correspond to $\rho^0$ production whilst the Rogers et al. curves \cite{Ref:Rogers} correspond to $J/\Psi$ data.}
\label{Fig:Dipole}
\end{figure}

The survival probability of small-sized dipoles in pQCD is close to 1 which can be contrasted strongly with that in dipole models where there are differences of up to a factor of 5, as shown in Figure \ref{Fig:Dipole}.  The HERA data on this is limited and this will be expanded at an EIC with measurements on nucleons and nuclei with much higher statistics.

\subsection{\textbf{How do fast probes interact with an extended gluonic medium?}}

\begin{figure}
\resizebox{0.5\textwidth}{!}{%
   \includegraphics{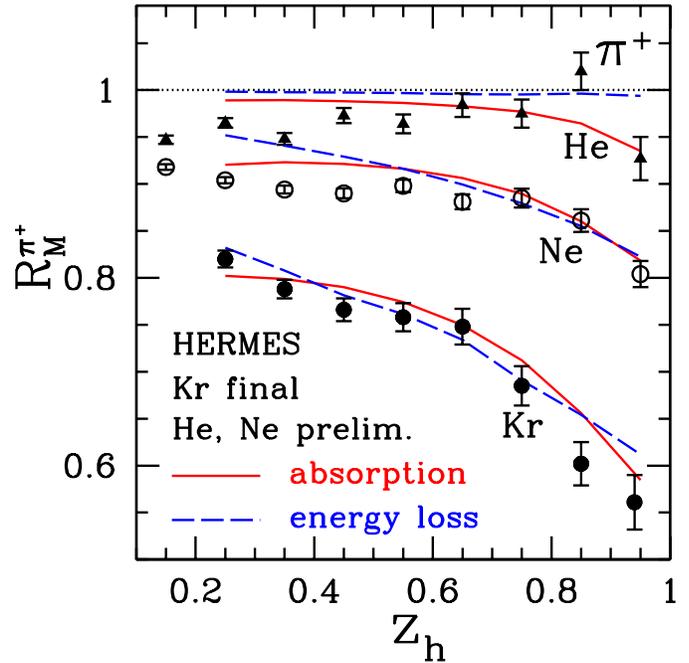}
}
\caption{The ratio of $\pi^+$ produced as a function of Z$_h$ for different nuclear species and compared to deuterium, where Z$_h$=E$_h$/$\mu$ is the fractional energy carried by a hadron (E$_h$) with respect to the virtual photon energy ($\mu$).  Also shown on the figure and compared to data are results from absorption and energy-loss models.  Compilation plot taken from \cite{Ref:Alberto}.}
\label{Fig:Had}
\end{figure}

In DIS on light nuclei, hadron suppression at higher transverse momentum has been observed which is analogous to, but smaller than, that observed for heavy A+A collisions at RHIC~\cite{Ref:JetQuenchingSTAR}.  Using nuclear DIS, one can study the details of the energy loss of particles traversing through ``cold" nuclear matter (as opposed to ``hot" nuclear matter in heavy-ion collisions, using the nucleons as femtometer-scale detectors.   This is important as, by determining the amount of energy loss in ``cold" nuclear matter, one is able to determine the correct amount of final-state energy loss in ``hot" nuclear matter.  This is one of the most important results to come from RHIC (see section~\ref{sec:Jets}).  The main question to be asked is what the time-scale is for the colour of the struck quark to be neutralised, acquiring a large inelastic cross-section for interaction with the medium.  Experimental data shows that the ratio of hadrons per nuclear DIS event as a function of virtual-photon energy ($\mu$) in heavier $e^+$+ A collisions is significantly reduced when compared to $e^+$+ deuterium~\cite{Ref:HERMES}.   In the absence of any nuclear effects, one would expect this ratio to be unity.  Both pre-hadron absorption and energy loss models have been applied to the HERMES data which is not of sufficient statistical significance to distinguish between them, as is shown in Figure \ref{Fig:Had}.  At an EIC, this measurement will be possible over a significantly much wider range of $\mu$ and crucially, is able to perform significantly significant measurements of charmed hadrons, which, as discussed in section~\ref{sec:Jets}, is of particular relevance to RHIC data.

\subsection{\textbf{What is the role of Pomerons (colour neutral excitations) in scattering off nuclei?}}

\begin{figure}
\vspace{0.5cm}
\resizebox{0.48\textwidth}{!}{%
   \includegraphics{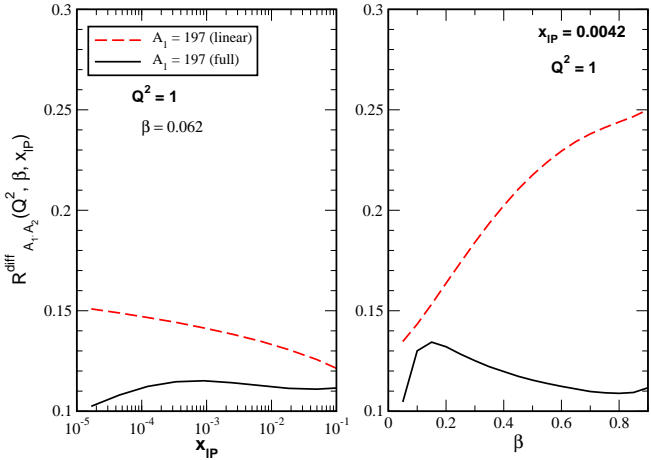}
}
\caption{The normalised ratio of diffractive structure functions in Au relative to deuterium nuclei, shown as a function of the variables $x_{\pom}$ and $\beta$ (where $\beta$ = \x/$x_{\pom}$) at a fixed \qs.  The curves show the expectations for different models of linear and non-linear QCD evolution effects which should be able to be differentiated at a future EIC.}
\label{Fig:Pomeron}
\end{figure}

The role of diffractive physics, which is the description for when an electron probe interacts with a colour-neutral object called a Pomeron (which can be visualised as a colourless combination of two or more gluons), is an important one in \ea physics as it is predicted to account for up to 30-40$\%$ of the total cross-section, as outlined in Figure \ref{Fig:Pomeron}.  There are two types of diffractive measurements: $i)$ coherent diffraction occurs when there is no nuclear break-up in the final state and $ii)$ incoherent diffraction occurs when the initial nucleus is not present in the final state.  Studies of coherent diffractive scattering are easier in a collider environment.  Performing these measurements at an EIC for the first time will allow us to directly probe the structure of the Pomeron and possibly identify its ``nature" and will provide for stringent tests on strong gluon field dynamics on QCD.

\subsubsection{Measuring Diffractive Events in \ea Collisions}

Diffractive physics is an important part of the \ea programme at an Electron-Ion Collider.  It is particularly useful as the differential cross-section, $d\sigma/dt \propto (xG(x,Q^2)^2$, where $t$ is the square of the difference between the final- and initial-state nuclei ($t = (p-p`)^2$ ).  Therefore, in order to make accurate measurements, it is desirable to measure precisely the value of $t$, by measuring both the initial- and final-state nuclei.  Measuring the final-state nuclei can be achieved through a Roman-Pot type detector, placed in the beam-pipe far upstream of the interaction.  However, it has been shown in Ultra-Peripheral Collisions by the STAR experiment that the value of $t$ required to break-up the nucleus (that is, change coherent diffraction to incoherent diffraction) is only approximately 30 MeV$^2$~\cite{Ref:STARUPC}.  This low value of $t$ is too small to kick the nucleus out of the beam and hence makes it un-measurable by a Roman Pot detector.  

One technique to overcome this obstacle is to measure the exclusive production of vector meson production.  In this scenario, it is possible to approximate the value of $t$ to that of $p_T^2$ of the vector meson.  One drawback of this technique is that it requires completely hermetic coverage and the event type is limited.

Other methods of measuring diffractive events exist, such as the Large Rapidity Gap method, which utilises one of the key signatures of diffractive events which is that there is a gap in the rapidity of the final state particles when compared to normal DIS events.  That is, an angular region in the direction of the scattered hadron/nucleus without particle flow.  Experimentally, the pseudorapidity ($\eta$) of the most forward particle in the detector is measured and compared to the maximum $\eta$ which can be measured.  At HERA it was estimated that in diffractive events, this gap would be 7.7 units wide.  However, in reality it was found that this gap was reduced to only 3 or 4 units by the spread coming from hadronisation.

\section{The connection to Heavy-Ion Collisions: the role of gluons at RHIC and the LHC}
\label{RHI}
Relativistic Heavy-Ion Collisions have provided a wealth of data over recent years on the search for a state of matter with partonic degrees of freedom.  Two of the major results on soft and hard probes are described below:

\subsection{Hadronic Flow}

The strong flow of hadrons has been observed in A+A collisions at RHIC, which, for the first time, is in agreement with hydrodynamical model predictions~\cite{Ref:Hydro}.  This is much greater than what can be produced in hadron-gas models and is indicative of the formation of a strongly-coupled medium.  These models suggest that the system produced in A+A collisions at RHIC is almost completely thermalised by 1 fm/$c$ after the initial collision.  However, the mechanisms by which rapid thermalisation is achieved are not fully understood as there is no information from QCD on thermalisation from first principles.  It is thought though that it is driven by low-x gluons with $k_T^2 < Q_S^2$, where $Q_S$ is the saturation scale discussed earlier.  The thermalisation time-scale in the hydrodynamic models is dependent upon the initial conditions used and hence, in order to understand these processes correctly requires knowledge of the momentum and spatial distributions of gluons in nuclei, $x$G$_A(x,Q^2,b)$.

\subsection{Quenching of jets and heavy flavour}
\label{sec:Jets}

The higher cross-sections and luminosities available for hard probes at RHIC (when compared to the AGS and SPS) has meant that precision studies have provided for a crucial role in understanding the matter formed.  The RHIC data has thrown up some surprising results.  One of the first hard-probe measurements was the attenuation of particles with high transverse momentum (p$_T$) in general and, more specifically, the disappearance of back-to-back jets in the most central collisions, indicating energy loss as they traversed through the strongly coupled dense matter~\cite{Ref:JetQuenchingSTAR}.  The initial gluon distributions play a crucial role in determining, quantitatively, the amount of energy lost.  The gluon distributions are strongly modified in nuclei (shadowing and saturation at low-\x and the EMC effect at medium-$x$).  As shown in Figure \ref{Fig:GluonDists} and discussed earlier, these gluon distributions are not well defined at lower \x and especially at values of \x which are relevant to forward physics at RHIC and mid-rapidity physics at the LHC.  Precision measurements of these quantities are required in order to utilise hard probes to diagnose the degrees of freedom of the matter produced at RHIC and the LHC.

An extra ability of the EIC will be in the charm sector.  Due to the ``dead cone effect", heavy (charm and bottom) hadrons are expected to lose less energy than the light hadrons as they cannot radiate gluons at small angles~\cite{Ref:DeadCone}.  Surprisingly, the data shows that this is not the case and the heavy mesons lose just as much energy as the light hadrons as they traverse the produced matter.  This is a challenging result for theory, requiring a re-assessment of the role of collisional energy loss and pre-hadron absorption in cold nuclear matter.  First results on this effect in $e$+A collisions for light hadrons has been shown for light nuclei by HERMES~\cite{Ref:HERMES}, but a study over a wide range of energies and A, especially for heavy flavours, still needs to be performed.

\section{The plans for an electron-ion machine}
\label{Sec:EIC}

The requirements for an \ea collider are driven by the region of \x and \qs space where saturation physics is applicable.  This region is depicted in Figure \ref{Fig:l+A} where it can be seen that the required region is at low \x and low $Q^2$.  In order to reach into this region, the following requirements are envisaged:

\begin{itemize}
\item Collisions of at least $\sqrt{s_{_{NN}}} > 60$\,GeV in order to go well beyond the range explored in the past fixed target $l$+A experiments outlined in Figure \ref{Fig:l+A}.
\item Luminosities of $L > 10^{30}$ cm$^{-2}$s$^{-1}$ are required to enable precise and definitive measurements of the gluon distributions of interest.
\item Provision of ion beams at different energies.  As shown at HERA,  this is mandatory in order to make direct measurements of  $F_L$, rather than inferring $F_L$ through scaling violations of $F_2$.
\item Provision of a wide range of ion species. Again, as shown in Figure \ref{Fig:l+A}, it is important to have high mass ions as these make the saturation region more accessible for lower energies. 
\end{itemize}

There are currently two proposals for the realisation of an Electron-Ion Collider in the USA.  Although the goals of the two machines are very similar, they are reached in different ways.  The next sections outline these two proposals followed by a section on current proposals for detectors for an EIC.

\begin{figure*}
\resizebox{0.5\textwidth}{!}{%
    \includegraphics{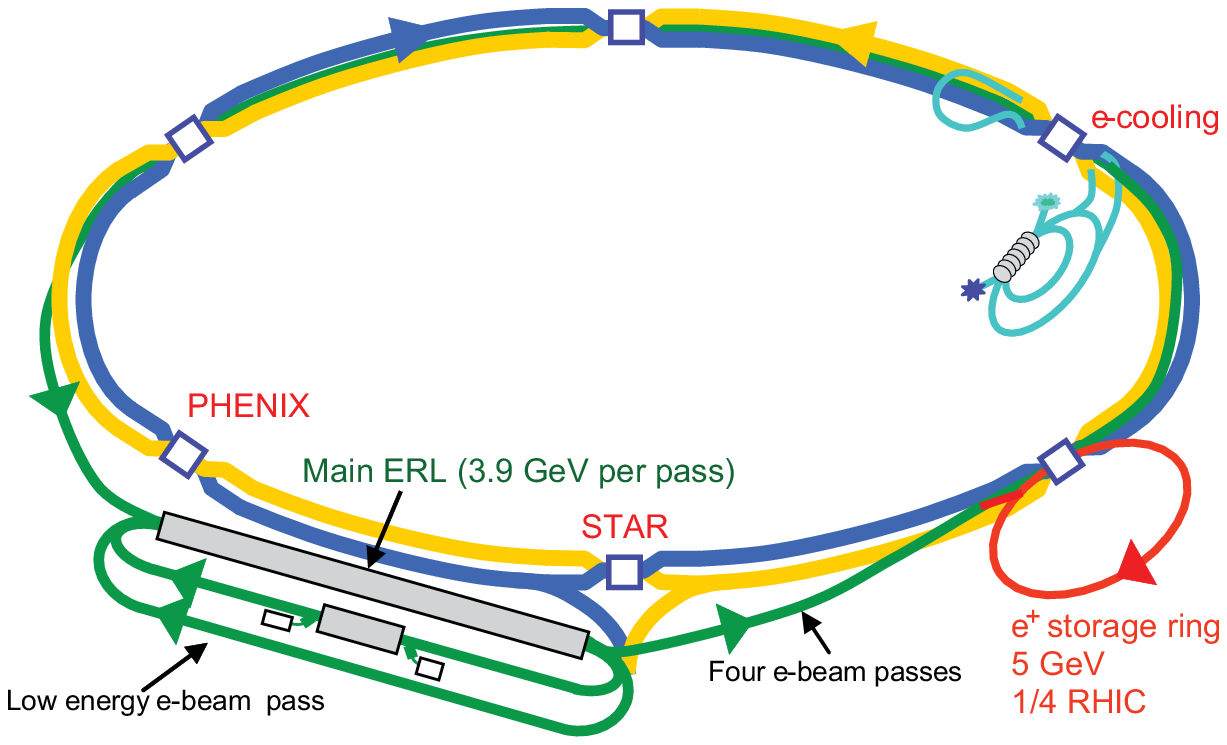}
}
 \resizebox{0.5\textwidth}{!}{%
    \includegraphics{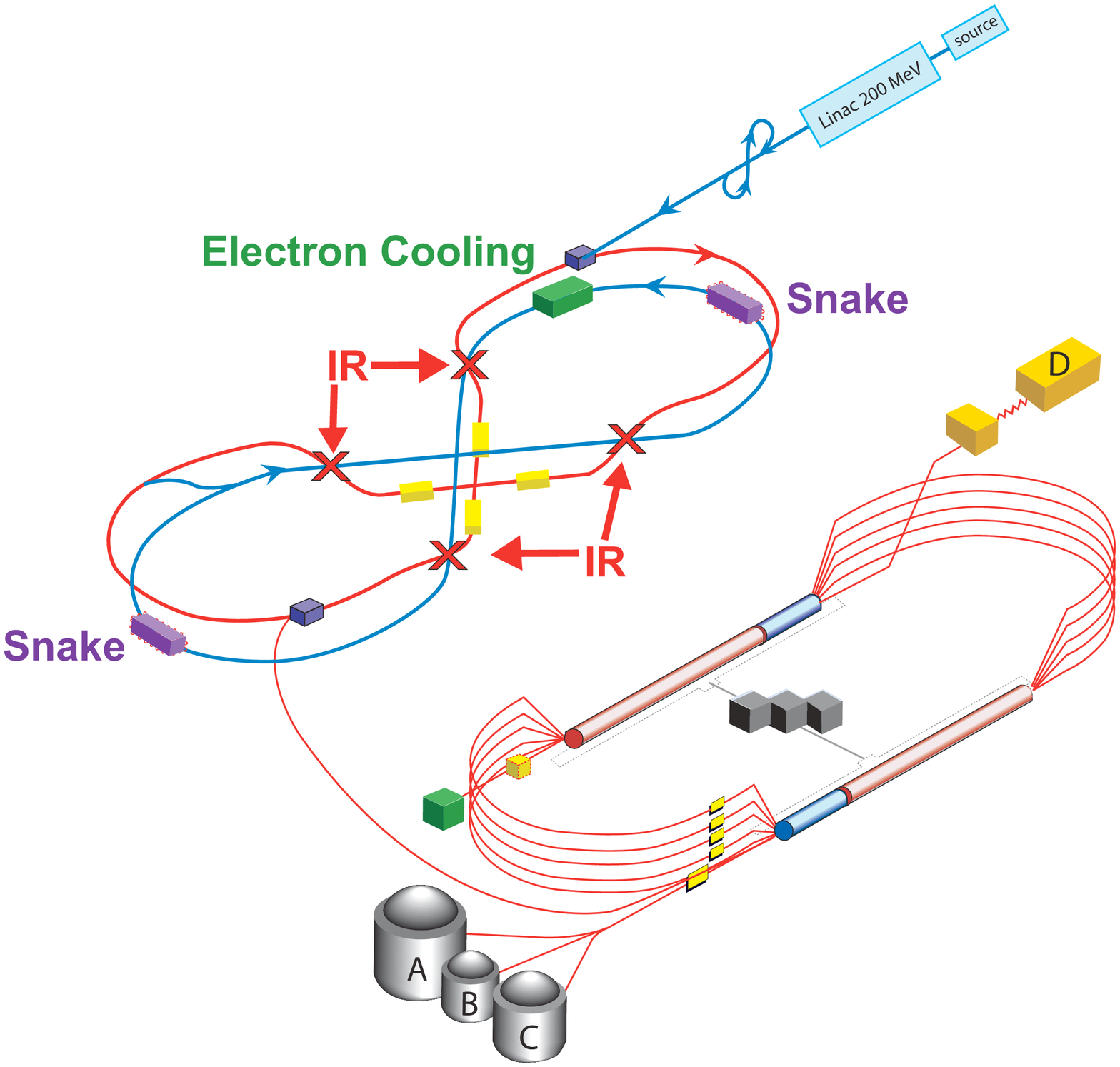}
}
 \vspace{3mm}
     \caption{Design layout of the eRHIC collider at BNL
         based on the Energy Recovery Linac (left)
         and the ELIC schematic layout (right) at the JLAB.}
    \label{fig:eic_concepts}
\end{figure*}

\subsection{ELIC at JLab}

At JLab, there currently exists an electron accelerator and they require an ion beam with which to collide it.  ELIC at JLab is envisioned to be an addition to the CEBAF accelerator, occurring after the currently planned upgrade to 12 GeV beams.  The CEBAF accelerator will be used as an injector into an electron storage ring, which will collide with a newly built ion complex.  The ion complex will be capable of accelerating polarised protons and unpolarised heavy ions.  The current design (outlined in the right panel of Figure \ref{fig:eic_concepts}) aims for a peak luminosity of 10$^{35}$ cm$^{-2}$s$^{-1}$ (well above the required luminosity) for collisions of 9 GeV electrons on 90 GeV/n ions, leading to $\sqrt(s_{_{NN}}) \sim 57$ GeV.

\subsection{eRHIC at BNL}

At BNL, there currently exists an A+A collider (RHIC).  It has capabilities to collide beams of polarised protons at energies up to 250 GeV and heavy-ions up to Au at energies up to 100 GeV/n.  The heavy-ion source will soon be upgraded to increase the capability to accelerate ions up to the mass of U.  Therefore, in order to realise an EIC at BNL, an electron accelerator is required to be built.  The option being pursued is outlined in the left panel of Figure \ref{fig:eic_concepts}, which shows the addition of a superconducting energy recovery linac (ERL) to the existing machine.  The current design aims for a peak luminosity of 10$^{34}$ cm$^{-2}$s$^{-1}$ (well above the required luminosity) for collisions of 30 GeV electrons on 130 GeV/n ions, leading to $\sqrt(S_{NN}) \gg 60$ GeV.  The additional 30$\%$ increase in energy can be achieved by not utilising the focusing magnets required for A+A collisions.

\subsubsection{Staged Proposal for eRHIC}

One recent proposal which is applicable at RHIC is for a staged approach to the construction of eRHIC.  This opportunity exists as the proposed upgraded RHIC luminosities  (referred to as RHIC II) have been achieved without the need for stochastic electron cooling.  The money saved, which was earmarked for this proposal, could be utilised in a staged approach.  This would have to be done in such a way as to maximise the accelerator capabilities for the available money and this, therefore, requires a minimisation of civil construction costs.  The other consideration is that, for this to be a real staged approach, that the design used be used in the full eRHIC.  The approach used is therefore a proposal to build the electron accelerator inside one of the currently unused interaction regions which exists at RHIC (IR 2).  Using this space, it is possible to build a 2 GeV ERL.  As this is a new proposal, the physics possibilities at this energy, for both polarised $e+p$ and \ea collisions are currently under investigation.  One advantage that this has is that, as it is cheaper and includes no civil construction, it can be realised at an earlier data than a full eRHIC.

\subsection{Detectors for an EIC}

One of the important tasks of a new collider is the design of new detectors as the current detectors at both JLab and RHIC are not suitable for \ea collisions.  Therefore, new detectors will have to be built and designed, taking into account the specific needs of both \ea and polarised $e+p$ physics.  That is, the ability to measure electrons which are scattered at low angles (as will occur in the \x and \qs region where saturation physics will be prevalent) as well as full hermetic coverage in order to measure, amongst other things, large rapidity gaps in diffractive.  Initial ideas for EIC detectors have been proposed and these are available in the literature~\cite{Ref:Caldwell}~\cite{Ref:Surrow}, though there is an ongoing effort to design the best detectors for the physics to be studied at an EIC.

\section{Summary}

In this paper, I have outlined the needs and requirements for an Electron-Ion Collider.  Precision measurements with an EIC will allow for the first time the exploration of the physics of the defining features of QCD, that is, gluons and their self-interactions.  These self-interactions are incredibly important and understanding them is essential to understanding the processes occurring in high energy heavy-ion collisions.  This physics can be studied at a high luminosity EIC (realised either at BNL (eRHIC) or at JLab (ELIC), which will provide access to the low $x,Q^2$ ``saturation" region, where the physics is dominated by gluons, as well as accessing the spin structure of gluons utilising the polarised proton beams.  As shown in Figure \ref{Fig:l+A}, this region is so far unexplored in current and previous experiments.  Whilst there has been significant progress in the realisation of an EIC and the development of a physics case, further R$\&$D on accelerator concepts such as high energy, high current ERLs and is essential.  This has been endorsed by the NSAC and a provision of US$\$$5 million per annum was recommended in the 2007 NSAC Long Range Plan.  

In conclusion, an EIC will provide unique capabilities for the study of QCD far beyond those available at existing facilities.  The EIC presents a unique opportunity in high energy nuclear and precision QCD physics.

%
%

\end{document}